\documentclass{sig-alt-release2}

\begin{document}
\conferenceinfo{WWW 2013 Companion,} {May 13--17, 2013, Rio de Janeiro, Brazil.}
\CopyrightYear{2013}
\crdata{978-1-4503-2038-2/13/05}
\clubpenalty=10000
\widowpenalty = 10000

\title{Popularity Prediction in Microblogging Network: A Case Study on Sina Weibo}

\numberofauthors{1}
\author{
\alignauthor
	Peng Bao, Hua-Wei Shen, Junming Huang, Xue-Qi Cheng\\
	\affaddr{Institute of Computing Technology, Chinese Academy of Sciences, Beijing, China}\\
	\email{pengbaocn@gmail.com, shenhuawei@ict.ac.cn, mail@junminghuang.com, cxq@ict.ac.cn}
}


\maketitle
\begin{abstract}
Predicting the popularity of content is important for both the host and users of social media sites. The challenge of this problem comes from the inequality of the popularity of content. Existing methods for popularity prediction are mainly based on the quality of content, the interface of social media site to highlight contents, and the collective behavior of users. However, little attention is paid to the structural characteristics of the networks spanned by early adopters, i.e., the users who view or forward the content in the early stage of content dissemination. In this paper, taking the Sina Weibo as a case, we empirically study whether structural characteristics can provide clues for the popularity of short messages. We find that the popularity of content is well reflected by the structural diversity of the early adopters. Experimental results demonstrate that the prediction accuracy is significantly improved by incorporating the factor of structural diversity into existing methods.
\end{abstract}


\section{Introduction}
Popularity prediction on social networks can help users sift through the vast stream of online contents and enable advertisers to maximize revenue through differential pricing for access to content or advertisement placement. Popularity prediction is challenging since numerous factors can affect the popularity of online content. Moreover, popularity is very asymmetric and broadly-distributed. Several pioneering work devoted to the characteristics and mechanisms of information diffusion~\cite{barabasi05, Wu2007PNAS, Yang2011WSDM}. Several efforts have been made to study the popularity prediction on social networks. Szabo et al.~\cite{Szabo2010CACM} found that the final popularity is reflected by the popularity in early period by investigating Digg and Youtube. A direct extrapolation method is then employed to predict the long-term popularity. Lerman et al.~\cite{Lerman2010WWW} modeled users' vote process on Digg by considering both the interestingness and the visibility of online content. Hong et al.\cite{Hong2011WWW} formulated the popularity prediction as a classification problem.

However, existing methods pay little attention to the structural characteristic of the propagation path of online content. In this paper, we consider the popularity prediction problem by studying the relationship between the popularity of online content and the structural characteristics of the underlying propagation network. The study is conducted on the Sina Weibo, the biggest microblogging network in China. Experimental results demonstrate that our method significantly outperforms the state-of-the-art method which neglects the structural characteristics of social networks. This indicates that the structural diversity would give us some insights to understand the mechanism of information diffusion and to predict the long-term popularity of a tweet.

\section{Problem Statement}
In this paper, the popularity prediction aims to predict the popularity $p(t_r)$ of a tweet at a \textit{reference time} $t_r$, given the forward information of this tweet before an \textit{indicating time} $t_i$. The indicating time $t_i$ is the time at which we observe the information of a tweet and the reference time $t_r$ is the time at which we predict the popularity of the tweet. The popularity $p(t)$ is measured by the number of times that a tweet is re-tweeted at time $t$.

\section{Findings and Methods}
We first study the structural characteristics of the forward path of tweets. Encouraged by the work in~\cite{Ugander12PNAS}, we investigate whether the final popularity of a tweet is well indicated by the structural characteristics of the network consisting of users that re-tweet the tweet at an earlier time. Specifically, we analyze the structural characteristics of a tweet with the following two measurements on its re-tweet path at $1$ hour after it is posted. The first measurement is \emph{link density}. Among all users that have forwarded the tweet $k$ at time $t_i$, link density is the ratio of the number of followship links to the number of all possible links. The other measurement is the \emph{diffusion depth}, which is the longest length of the path from the submitter to any user that has retweeted the tweet $k$ at time $t_i$.

\begin{figure}
\vspace{-15pt}
\centering
	\includegraphics[height=1.4in, width=1.6in]{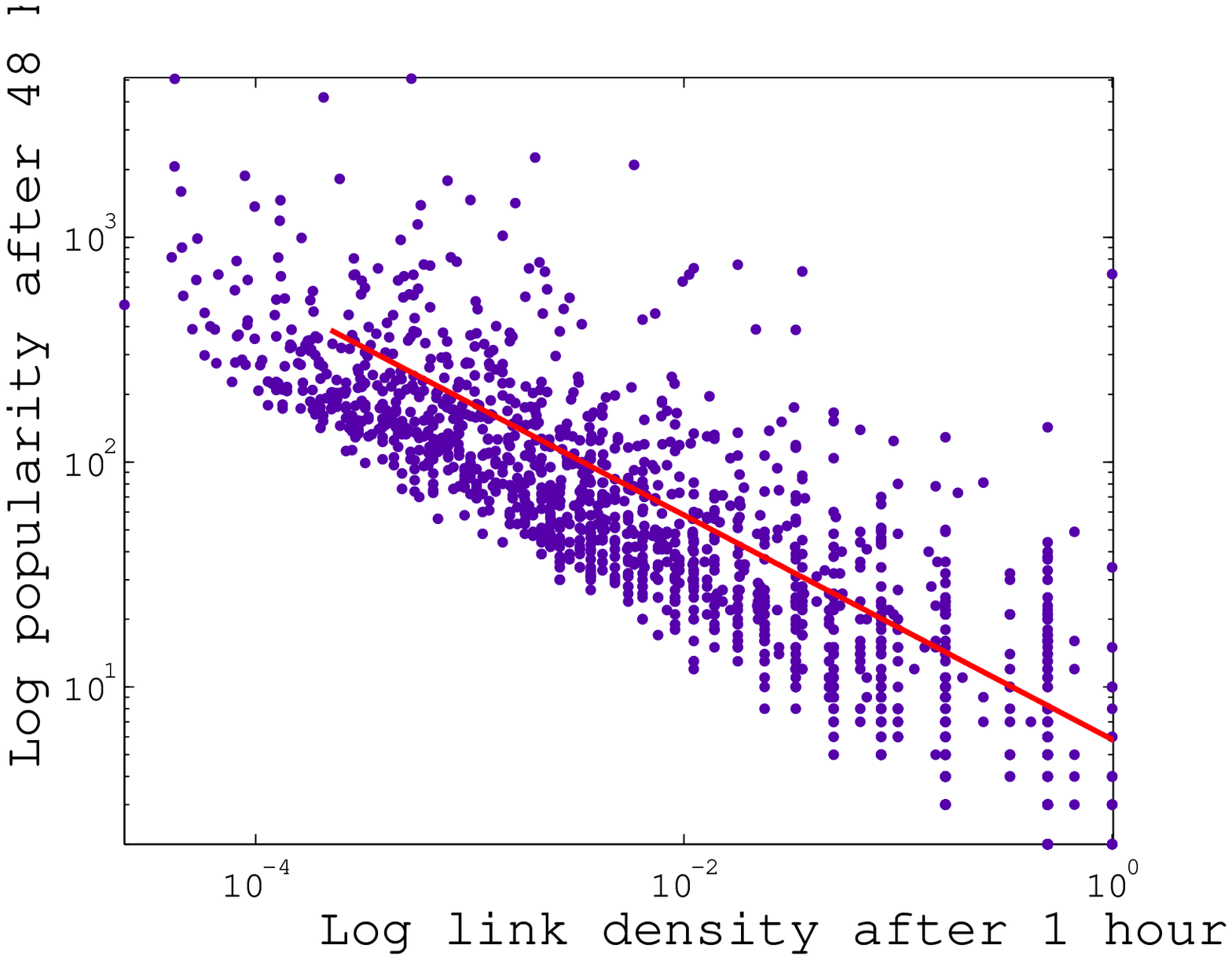}
	\includegraphics[height=1.4in, width=1.6in]{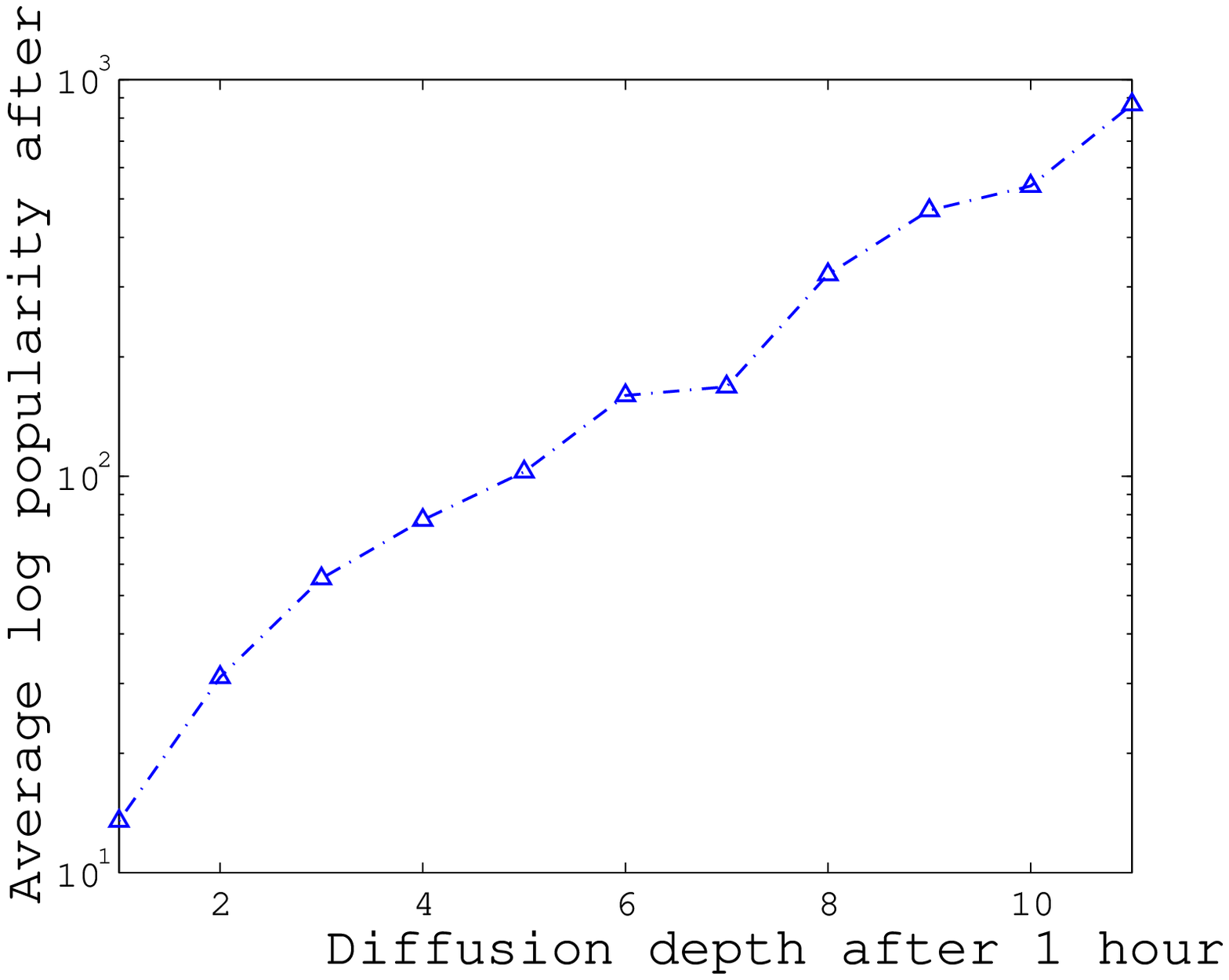}
\caption{Structural characteristics}
\label{fig:structural}
\vspace{-15pt}
\end{figure}

We report the final popularity of a tweet with respect to the link density and the diffusion depth. As shown in Figure~\ref{fig:structural}, there exists a strong negative linear correlation between the final popularity and the link density, and there exists a strong positive near-linear correlation between the final popularity and the diffusion depth. This finding tells us that a diverse group of earlier users, reflected with low link density and large diffusion depth, leads to a wide spreading of a tweet. Therefore, the structural characteristics of diffusion paths of a tweet at an earlier time can help predict its final popularity.

Based on the above findings, we propose two improved approaches to predict the final popularity using earlier popularity and structural characteristics. We estimate the logarithmic final popularity with a combination of the logarithmic early popularity and the logarithmic link density,
\begin{equation}\label{eq:LRLD}
\ln{\hat{p}_k(t_r)} = \alpha_1 \ln{p_k(t_i)} + \alpha_2 \ln \rho_k(t_i) + \alpha_3,
\end{equation}
where $\rho_k(t_i)$ is the link density at or before time $t_i$, and $\alpha_1$, $\alpha_2$ and $\alpha_3$ are global coefficients that will be learned from the data. Similarly, we define a diffusion depth version to estimate the logarithmic final popularity of a tweet as
\begin{equation}\label{eq:LRDD}
\ln{\hat{p}_k(t_r)} = \beta_1 \ln{p_k(t_i)} + \beta_2 d_k(t_i) + \beta_3,
\end{equation}
where $d_k(t_i)$ is the diffusion depth of the tweet $k$ at or before time $t_i$, and $\beta_1$, $\beta_2$ and $\beta_3$ are also global coefficients.

To demonstrate the effectiveness of our proposed approaches, we compare them with a baseline approach which estimates the final popularity with the early popularity alone~\cite{Szabo2010CACM}. The baseline predicts the final popularity using
\begin{equation}\label{eq:LR}
\ln{\hat{p}_k(t_r)} = \gamma_1\ln{p_k(t_i)} + \gamma_2,
\end{equation}
where $\gamma_1$ and $\gamma_2$ are also global coefficients that will be learned from the data.

\section{Experiments}

We use Sina Weibo dataset published by WISE 2012
Challenge\footnote{http://www.wise2012.cs.ucy.ac.cy/challenge.html}. We select the tweets posted during July 1-31, 2011 and all the re-tweet paths occurred during July 1-August 31, 2011. The data set consists of 16.6 million tweets. This data set also contains a snapshot of the social network of Sina Weibo. The social network contains 58.6 millions of registered users and 265.5 millions of following relations.

\begin{table}
\centering \caption{Prediction error of three approaches.}\label{table:1}
\begin{tabular}{|c|c|c|}
\hline
Primitive type & RMSE & MAE\\\hline
Baseline & 0.77 & 0.57\\\hline
with link density & 0.63 & 0.45\\\hline
with diffusion depth & \textbf{0.61} & \textbf{0.43}\\\hline
\end{tabular}
\end{table}

We take $75\%$ of all the tweets in the dataset as the training set and the rest $25\%$ as the
testing set. The predictions are evaluated with \emph{RMSE} (root mean squared
error) and \emph{MAE} (mean absolute error). As reported in Table \ref{table:1},
the approach incorporating the link density significantly reduces the prediction error compared with
the baseline, and the approach incorporating the diffusion depth performs even better. Here, the
values of $\alpha_2$ and $\beta_2$ in previous formulas that we learned from the data is 0.04 and 0.07 separately.

The results empirically demonstrate that early structural characteristics affect the
final popularity. Low link density and long diffusion path implies that a
tweet is more probably spread to different parts of the network, which helps the tweet
become known by a greater population.

\section{Conclusions}
In this paper, we have studied how to predict the popularity of short message in Sina Weibo. We find that structural characteristics provide strong evidence for the final popularity. A low link density and a deep diffusion usually lead to wide spreading, capturing the intuition that a diverse group of individuals spread a message to wider audience than a dense group. Based on such a finding, we propose two approaches by incorporating the early popularity with the link density and the diffusion depth of early adopters. Experiments demonstrate that the proposed approaches significantly reduce the error of popularity prediction. Our finding provides a new perspective to understand the popularity prediction problem and is helpful to build accurate prediction models in the future.

\section{Acknowledgements}
This work is funded by the National Natural Scientific Foundation of China under grant Nos. 61232010, 61202215 and National Basic Research Program of China (the 973 program) under grant No. 2013CB329602. This work is partly funded by the Beijing Natural Scientific Foundation of China under grant No. 4122077. This work is also supported by Key Lab of Information Network Security, Ministry of Public Security.

\end{document}